\documentclass[aps,pra,twocolumn,groupedaddress, showpacs]{revtex4}

\usepackage{graphicx}
\usepackage[dvips]{color}

\newcommand{\ket}[1]{|#1\rangle}

\begin{document}


\title{A pulsed Sagnac source of narrowband polarization-entangled photons}

\author{Onur Kuzucu}
\author{Franco N. C. Wong}
\affiliation{Research Laboratory of Electronics, Massachusetts Institute of
Technology, Cambridge, Massachusetts 02139, USA}


\date{\today}

\begin{abstract}

We demonstrate pulsed operation of a bidirectionally pumped polarization Sagnac
interferometric down-conversion source and its generation of narrowband,
high-visibility polarization-entangled photons.  Driven by a narrowband,
mode-locked pump at 390.35\,nm, the phase-stable Sagnac source with a type-II
phase-matched periodically poled KTiOPO$_4$ crystal is capable of producing 0.01
entangled pair per pulse in a 0.15-nm bandwidth centered at 780.7\,nm with 1\,mW
of average pump power at a repetition rate of 31.1 MHz.  We have achieved a mean
photon-pair generation rate of as high as 0.7 pair per pulse, at which
multi-pair events dominate and significantly reduce the two-photon
quantum-interference visibility.   For low generation probability $\alpha$, the
reduced visibility $V=1-\alpha$ is independent of the throughput efficiency and
of the polarization analysis basis, which can be utilized to yield an accurate
estimate of the generation rate $\alpha$.  At low $\alpha$ we have characterized
the source entanglement quality in three different ways: average
quantum-interference visibility of 99\%, the Clauser-Horne-Shimony-Holt $S$
parameter of $2.739 \pm 0.119$, and quantum state tomography with 98.85\%
singlet-state fidelity. The narrowband pulsed Sagnac source of entangled photons
is suitable for use in quantum information processing applications such as
free-space quantum key distribution.
\end{abstract}


\maketitle


\section{Introduction}

Photonic entanglement is an essential resource for many quantum information
processing applications, such as linear-optics quantum computing
(LOQC) \cite{klm01}, quantum teleportation \cite{teleport}, and quantum key
distribution (QKD) \cite{Ekert91}.  The most widely used method for generating
polarization-entangled photon pairs is spontaneous parametric down-conversion
(SPDC), which can be realized in a number of configurations and by using
different crystals \cite{kwiat, fiorentino04, kwiat05, kim06, wong06, zeilinger,
yhkim, hodelin}.  A well optimized SPDC source with high flux and high
entanglement quality can be utilized to develop advanced enabling quantum
technology such as a random number generator \cite{fiorentino}.  Certain
applications such as quantum communications and LOQC are often designed to
operate with a system clock to allow for synchronization and, for that purpose,
a pulsed source of entangled photons is needed.  In other applications such as
free-space entanglement-based QKD in the daylight, pulsed operation may be
preferred to simplify signaling of the arrival times of the entangled photons
and to provide temporal discrimination against undesirable background light.
Many SPDC sources, however, are continuous-wave (cw) pumped which cannot yield
any timing information about when an entangled photon pair is generated.  It is
therefore of interest to develop an efficient pulsed source of entangled photons
that can offer unique operational and application-specific advantages over a cw
SPDC source.

We have previously developed a cw SPDC source of polarization-entangled photons
using a polarization Sagnac interferometer (PSI) configuration that is highly
efficient and yields a high visibility in two-photon quantum interference
\cite{kim06,wong06}. The Sagnac source configuration is a phase-stable,
single-crystal implementation of interferometrically combined outputs of two
identical, coherently-driven down-converters \cite{kwiat94,shapiro00}.  The PSI
configuration eliminates the need for spatial (aperture), spectral (interference
filter), and temporal (timing compensator) filtering because the two
down-converter outputs are completely indistinguishable \cite{fiorentino04}. All
the output photons are strongly polarization-entangled without the need for
filtering, thus leading to a much higher generation efficiency than that of
other approaches.

In principle, the Sagnac configuration can be used for both cw and pulsed operation.  Shi and Tomita have previously utilized a non-polarizing
Sagnac interferometer to realize a highly stable pulsed down-conversion entanglement source \cite{shi}.  The use of a non-polarizing beam
splitter incurs a 50\% reduction in the entanglement generation rate because when both photons exit the same port of the beam splitter (half of
the time) it does not produce polarization entanglement.  In this work, we employ a polarizing beam splitter (PBS) to eliminate this 50\% loss,
and  we demonstrate the first pulsed operation of a Sagnac down-conversion source pumped by a narrowband, mode-locked ultraviolet (UV) source.
Unlike most pulsed SPDC sources that are pumped by femtosecond lasers \cite{yhkim,hodelin}, the pulsed Sagnac source is pumped with $\sim$50-ps
pulses with a narrow spectral bandwidth. While the cw and pulsed Sagnac sources share a common design, there can be major differences between
the two types of sources.  In cw down-conversion the output flux is often limited by available cw pump power, especially in the UV region.
Therefore cw sources are often characterized by a normalized generation efficiency in terms of the number of entangled pairs generated per
second per nanometer of detection bandwidth for 1 mW of pump power.  In general, the situation is different for pulsed pumping because high peak
power is easily obtainable at many wavelengths, either directly from a laser or by efficient harmonic generation of a high power pulsed laser
\cite{kuzucu07}. The high peak pump power for pulsed down-conversion benefits applications that can take advantage of a substantial pair
generation probability per pulse such as those requiring multiphoton coincidences for the generation of novel multipartite states in LOQC
\cite{kiesel, eibl}.  On the other hand, with high generation probabilities one must be careful with the impact of multiple-pair events on the
system performance of some quantum information processing applications. For instance, in free-space entanglement-based QKD the entanglement
source may be strongly driven to achieve a desirable key generation rate but it should not exceed a level that may compromise the security of
the generated secret keys \cite{ma}.

We have developed a pulsed Sagnac source based on type-II phase-matched
periodically poled KTiOPO$_4$ (PPKTP) and pumped with a custom-built
high-power mode-locked UV source at 390.35\,nm \cite{kuzucu07}. The Sagnac
source is capable of producing polarization-entangled photon pairs at
780.7\,nm with a mean pair generation of as much as one per pulse, even in a
small bandwidth of 0.15\,nm.  The narrowband pulsed Sagnac source is
particularly suitable for free-space entanglement-based QKD\@. Pulsed
operation affords synchronized detection of the entangled photons and
provides temporal discrimination against temporally random background photons or
detector dark counts.  The narrowband outputs can be used with narrowband
spectral filtering that is essential for blocking out ambient light during
daylight operation.

In the next section, we describe our experimental setup in the construction of
the pulsed Sagnac source.  In section III, we characterize the entanglement
quality of the PSI output with three different types of measurements: two-photon
quantum interference, violation of Clauser-Horne-Shimony-Holt (CHSH) form of
Bell's inequality \cite{CHSH}, and quantum state tomography.  In section IV, we
focus on the strongly pumped Sagnac source with its high-flux outputs and the
accompanying visibility degradation due to occurrence of multiple-pair events,
before concluding in section V.

\section{Experimental Setup}

We develop the narrowband pulsed Sagnac source for potential use in free-space
entanglement-based QKD that must satisfy a number of operational constraints.
For line-of-sight free-space QKD it has been suggested that the optimal
operating wavelength is around 780\,nm that takes into account atmospheric
transmission and the characteristics of commercially available Si avalanche
photodiodes (APDs) as single-photon counters \cite{nordholt02}.  Commercially
available Si APDs typically have a detection quantum efficiency of $\sim$50\%
and a rise time of $\sim$300\,ps.  The pump for the Sagnac source should have a
pulse width that is small compared with the detector rise time so that the pump
does not add to the timing uncertainty. At the same time, we prefer to
operate the down-converter in a quasi-cw manner with a pump bandwidth that is
small compared with the down-conversion phase-matching bandwidth or the spectral
filter bandwidth. Otherwise, some of the photon pairs generated by the broadband
pump would be outside of the measurement bandwidth and the system throughput
efficiency is reduced.  We choose the pump requirements to have a pulse width of
100\,ps or less and a bandwidth of less than 0.1\,nm. A narrowband
down-conversion output allows the use of narrowband spectral filters inserted
before the Si APD detectors to screen out background light from the sky during
daylight QKD operation. The pulse width and bandwidth requirements can
accommodate a range of pump operating conditions and we note that transform
limited pulses are not required.

Figure 1 shows the experimental setup for the pulsed Sagnac source together with
the measurement apparatus.  We will briefly describe the PSI configuration and
more details can be found in previous cw implementations
\cite{kim06,wong06,zeilinger}.  The PSI, as shown in the dashed box of Fig.~1,
is composed of two flat mirrors and a PBS that serves as the input and output
optical element.  Within the interferometer is a nonlinear crystal for
generating entangled photons by SPDC\@.  Bidirectional pumping of the crystal in
counter-propagating directions creates coherent superposition of the
counter-propagating down-conversion outputs at the PBS, whose output photons are
entangled in polarization. Interferometric combination of two outputs is usually
sensitive to path-length perturbations, but the PSI configuration eliminates the
need for path-length stabilization through its common path arrangement
\cite{kim06}. Moreover, the phase of the output state can be chosen by simply
adjusting the relative phase between the horizontally ($H$) and vertically ($V$)
polarized components of the pump.

For our pulsed Sagnac source, we used a 10-mm long flux-grown PPKTP crystal from
Raicol with a grating period of 7.85\,$\mu$m for type-II phase matching and
degenerate wavelength output at 780.7\,nm.  The crystal was temperature
stabilized at a phase-matching temperature of 28.6\,$^\circ$C using a
thermoelectric heater with a temperature stability of better than
0.1\,$^\circ$C\@.  Using cw second-harmonic generation, we measured an effective
nonlinear coefficient, $d_{\rm eff} = (2/\pi)d_{\rm 24}$ of $\sim$2.4\,pm/V\@.
To ensure that the counter-propagating pump components reach the crystal at the
same time, thus eliminating any temporal distinguishability between the two
outputs, we positioned the crystal at the center of the interferometer within
$\sim$1\,mm, which is much less than that spanned by the 50-ps pump pulse.

The pulsed UV pump centered at 390.35\,nm was a compact, home-built narrowband
picosecond UV source based on frequency quadrupling of an amplified mode-locked
fiber laser \cite{kuzucu07}. It had a bandwidth of less than 0.1\,nm, full-width
at half maximum (FWHM).  The passively mode-locked erbium-doped fiber laser with
a narrowband intracavity filter was operated at 1561.4\,nm at a repetition rate
of 31.1\,MHz.  The fiber laser output was amplified with a 5-W
polarization-maintaining erbium-doped fiber amplifier and followed by two stages
of single-pass second-harmonic generation to yield the pulsed UV output with a
maximum average power of 400\,mW and a pulse width of $\sim$50\,ps.  Because of
the frequency quadrupling configuration, there was a significant amount of
residual 780.7\,nm light in the UV source output. As a UV pump for the Sagnac
source, it was necessary to filter this undesirable near-infrared (near-IR)
light before reaching the PSI\@. We used a pair of dichroic mirrors (DMs) that
were highly transmissive at 780.7\,nm and highly reflective at 390.35\,nm, and
constructed a 7-bounce passage for the pump beam to filter the residual
780.7\,nm light. We estimate that a suppression of 140\,dB was achieved using
this multiple-bounce dichroic mirror pair arrangement with  negligible loss of
UV power.

The linearly polarized, collimated UV beam after the DM filter was passed
through a half-wave plate (HWP) and a quarter-wave plate (QWP) to adjust the
relative phase and amplitude between the $H$- and $V$-polarized components. We
focused the pump into the PSI using a plano-convex lens with a 15-cm focal
length and with broadband anti-reflection (AR) coating that covered both UV and
near-IR wavelengths. The PPKTP crystal was centered at the pump focal spot that
was measured to have a beam diameter of $\sim$90\,$\mu$m.  The Sagnac
configuration calls for an input/output PBS that works at both the pump and
down-conversion wavelengths.  Instead, we used a commercially available standard
PBS cube designed and AR-coated for 780\,nm but not for 390\,nm.  The PBS
provided acceptable extinction ratio ($\sim$20:1) to separate the $H$ and $V$
components of the UV pump, with absorption and reflection losses of 35\% at
390\,nm.  The PBS served to produce counter-propagating pumps for
bidirectional pumping of the down-conversion crystal.  An AR-coated
dual-wavelength half-wave plate (DW-HWP) designed for 390.35\,nm and
780.7\,nm rotated the $V$-polarized pump to $H$ polarization that was required
(along the crystal's $y$-axis) for type-II phase matching. We note that the low
extinction ratio PBS introduced a small amount of $V$-polarized pump at the
crystal but it was not phase matched to generate any down-converted photon
pairs.  Before the signal and idler outputs from the two bidirectionally pumped
down-converters were combined at the PBS, the clockwise circulating output was
rotated by 90$^\circ$ with the DW-HWP such that the signal and idler beams after
the PBS were spatially separated, as indicated in Fig.~1.

Each of the signal and idler outputs was collimated with a 15-cm focal length
plano-convex lens and passed through a matching circular aperture for
spatial-mode filtering. We sent the output beams to polarization analyzers
each comprising a HWP and a high-extinction polarizer to measure the
polarization states. A QWP could also be added for quantum state
tomography measurements. We filtered both beams spectrally using
interference-filters (IFs) with a FWHM bandwidth of $\sim$0.15\,nm centered at
780.7\,nm and with peak transmission of 60\%. The outputs were subsequently
focused onto two PerkinElmer Si APDs with $\sim$50\% detection efficiency. We
recorded individual APD counts as well as coincidence counts and a 1.8-ns
coincidence window was used \cite{coin_det}.

\begin{figure}[htb]\label{figure1}
\centerline{\includegraphics[width=8cm]{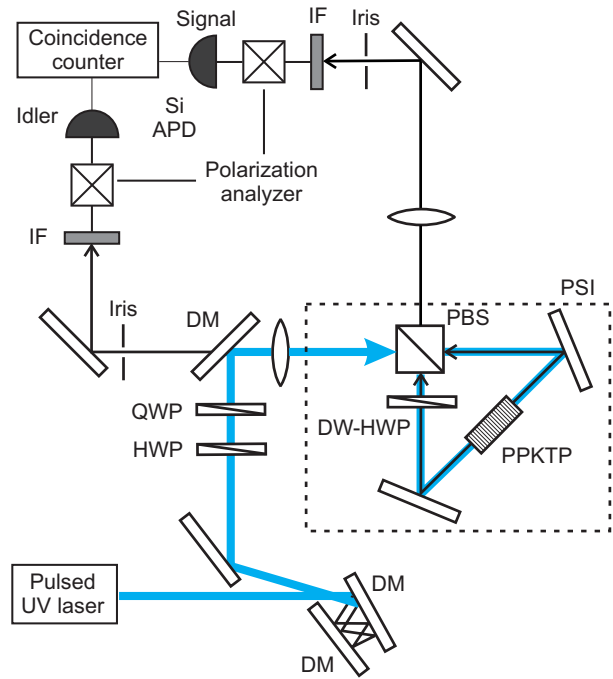}} \caption{(Color
online) Experimental setup showing the bidirectionally pumped polarization
Sagnac interferometer (PSI) in the dashed box.  The generated
polarization-entangled signal and idler outputs are analyzed in coincidence
measurements under different operating conditions.  IF, interference filter;
PBS, polarizing beam splitter; DM, dichroic mirror; HWP, half-wave plate; QWP
quarter-wave plate; DW-HWP, dual-wavelength half-wave plate.}
\end{figure}

We aligned the Sagnac interferometer such that the two down-converter outputs
for the signal (and for the idler) were spatially and temporally
overlapped at the PBS output to erase any "which-path" information. This
required symmetric positioning of the PPKTP crystal inside the Sagnac
interferometer. At this point, the pair generation rates from both paths were
balanced and the necessary input pump polarization adjustments were made to
generate a singlet-state output
\begin{equation}
\ket{\psi^-} = (\ket{H_S V_I} - \ket{V_S H_I})/\sqrt{2}\,, \label{singlet}
\end{equation}
where the subscript $S$ ($I$) refers to the signal (idler) photon.  In the next
two sections, we show the results of our source characterization in the low and
high flux regimes.

\section{Low-flux characteristics}
Unlike most cw SPDC sources, the combination of a high-power pulsed UV pump and
the efficient PSI source can lead to a substantial pair generation probability
per pulse.  In order to assess the entanglement quality of the pulsed Sagnac
source with minimal degradation due to multiple-pair events, we measured the
characteristics of the PSI output at low pumping powers.  In particular, we made
measurements of the two-photon quantum interference, the CHSH form of Bell's
inequality, and quantum state tomography of the PSI output state.  We show that
the three types of characterization methods are consistent with each other.

\subsection{Quantum interference visibility}

The signal-idler quantum interference measurement in two incompatible
polarization bases, $H$-$V$ and $\pm$45$^\circ$ antidiagonal-diagonal
($A$-$D$), is a common and relatively simple method to assess the entanglement
quality of the source.  For each of the polarization measurement bases, the
signal polarization analyzer angle is set at one of the basis polarization axes
and the coincidence counts are monitored as a function of the idler analyzer
angle.  The quantum-interference visibility is given by
\begin{equation}
V = \frac{C_{\rm max} - C_{\rm min}}{ C_{\rm max} + C_{\rm min} }\,, \label{V}
\end{equation}
where $C_{\rm max}$ and $C_{\rm min}$ are the maximum and minimum coincidence
counts, respectively, and an ideal entanglement source yields $V=1$.

\begin{figure}[thb]\label{figure2}
\centerline{\includegraphics[width=7cm]{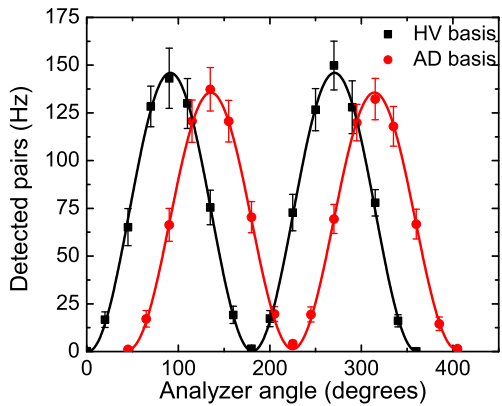}}
\caption{(Color online) Quantum interference measurements for input pump of
0.1\,mW and  sinusoidal fits in the $H$-$V$ (squares) and $A$-$D$ (circles)
bases. $H$-$V$ ($A$-$D$) visibility is 99.79\%
(98.11\%).}
\end{figure}

We observed the quantum interference fringes at different pump power levels and
different aperture sizes. Figure~2 shows the coincidence counts in the $H$-$V$
and $A$-$D$ bases and the corresponding sinusoidal fits to data at a pump power
of 0.1\,mW, as measured at the entrance to the PSI\@.  Each data point
represents an average of 30 1-s measurements, and we apply no background-count
subtraction to the coincidence count data in Fig.~2. Using the measured maximum
and minimum coincidence counts we obtain from Eq.~(2) an interference visibility
of $99.79\pm 0.38$\% in the $H$-$V$ basis and $98.11\pm 1.16$\% in the $A$-$D$
basis. The data in Fig.~2 were taken using a circular aperture with a full
collection divergence angle of $\sim$13\,mrad.  From the singles count rate of
$\sim$1,600/s we infer a $\sim$9.5\% conditional detection probability, limited
in large part by the Si APD detector efficiency of $\sim$50\%, the IF
transmission of $\sim$36\% (60\% peak transmission and 60\% double-Lorentzian
filter shape), and the measured 95\% transmission efficiency through the other
optical components.  The remaining reduction in the conditional detection
probability can be attributed to spatial filtering that was provided by the 2-mm
diameter aperture, which is a typical problem due to the multimode nature of
SPDC with an unfocused or weakly focused pump \cite{zeilinger}.  We measured a
Si APD dark-count rate of $\sim$50/s, and the non-parametric fluorescence counts
were less than 5\% of the measured singles which, at this pump level, did not
contribute to any accidental coincidences.

Figure~3 shows the quantum interference measurements at a pump power of 1.1\,mW
and with all other parameters being the same as in Fig.~2.  The singles rate
increased to 16,000\,s$^{-1}$ at this power level, and the quantum-interference
visibilities are found to be $98.04\pm 0.35$\% in the $H$-$V$ basis and
$96.64\pm 0.46$\% in the $A$-$D$ basis. From the singles count rate and the
9.5\% conditional detection probability, one can estimate the pair generation
rate to be 1.1\% per pulse.   This high pump power caused the slight reduction
in visibilities due to multiple-pair emission events, which are analyzed in
detail in the next section.

\begin{figure}[thb]\label{figure3}
\centerline{\includegraphics[width=7cm]{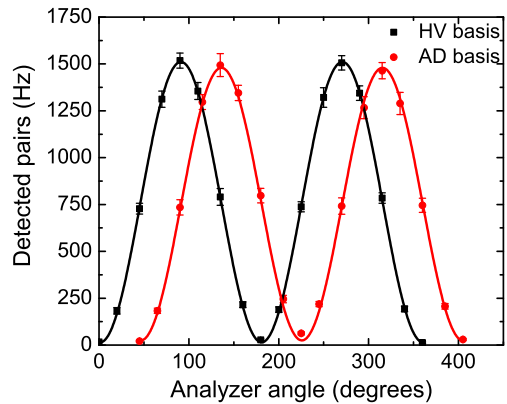}}
\caption{(Color online) Quantum interference measurements for input pump of
1.1\,mW and sinusoidal fits in the $H$-$V$ (squares) and $A$-$D$ (circles)
bases. With slightly higher photon pair generation rates, we obtain
98.04\% $H$-$V$ visibility and 96.64\% $A$-$D$ visibility.}
\end{figure}

We have also measured the variation in the quantum-interference visibility at
different divergence angles, as set by the aperture diameters, and
at a constant  input power of 1.1\,mW\@.  For each data point in Fig.~4, we
average 30 1-s measurements of $C_{\rm max}$ and $C_{\rm min}$, and
obtain the visibility calculated from Eq.~(\ref{V}). At a given pump power a
larger aperture size allowed more light to be collected which increased the
effective pair generation rate.   As a result, multiple-pair events increased
(see next section for details) and caused the visibility to degrade slightly, as
shown in Figure~4 for the measurements in the $H$-$V$ (squares) basis. The
$A$-$D$ data shows a more pronounced deterioration in the visibility at larger
collection angles.  The $H$-$V$ basis was the natural basis aligned with the
PPKTP's crystal principal axes.  Therefore, measurements in the $A$-$D$ basis
required the coherent superposition of the two counter-propagating
down-converter outputs, and the entanglement quality was sensitive to their
spatial mode distinguishability and to the phase variation of the output state
across the spatial extent.  In our case, the $A$-$D$ data in Fig.~4 clearly show
that the singlet-state entangled output of the PSI degraded at larger divergence
angles.  In designing the Sagnac source for a specific application, one must
therefore consider the trade-off between generation efficiencies and
entanglement quality as a function of the collection angles.

\begin{figure}[bh]\label{figure4}
\centerline{\includegraphics[width=7cm]{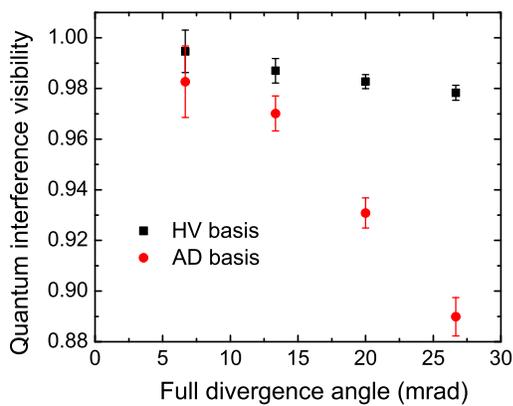}} \caption{(Color
online) Quantum interference visibilities in the $H$-$V$ (squares) and $A$-$D$
(circles) bases for various aperture sizes at a constant input pump power of
1.1\,mW.}
\end{figure}

\subsection{CHSH S-parameter measurements}

Violation of the CHSH form of Bell's inequality \cite{CHSH} is another common
method of entanglement characterization by measuring its $S$ parameter.  For an
ideal entangled state a maximum value of $S=2\sqrt{2}$ is predicted by quantum
mechanics, whereas $S$ cannot be greater than 2 classically.  Therefore, the
state is considered nonclassical if $2<S\le 2\sqrt{2}$ and the closer $S$ is to
$2\sqrt{2}$, the higher is the entanglement quality.  We follow the measurement
procedure of Ref.~\cite{kim06}.  The $S$-parameter is given by four measurement
expectation values \cite{CHSH}
\begin{equation}
S=\left|E(\theta_S,\theta_I) + E(\theta_S,\theta_I') - E(\theta_S',\theta_I) +
E(\theta_S',\theta_I')\right|\,, \label{S}
\end{equation}
where $\theta_S$ and $\theta_I$ are the polarization analyzer angles for signal
and idler, respectively.  We chose to maximize $S$ by using $\theta_S = -\pi/4$,
$\theta_S' = 0$, $\theta_I = 5\pi/8$, and $\theta_I' = 7\pi/8$.  Each
expectation value $E(\theta_S, \theta_I)$ is obtained from coincidence
measurements in 4 different polarization analyzer combinations: the chosen
signal and idler set and their orthogonal sets: $(\theta_S, \theta_I)$,
$(\theta_S,\theta_I+\pi/2)$, $(\theta_S+\pi/2,\theta_I)$ and
$(\theta_S+\pi/2,\theta_I+\pi/2)$.  One calculates the expectation value
according to
\begin{equation}
E(\theta_S, \theta_I) =\frac{C_{++}-C_{+-}-C_{-+}+C_{--}}{C_{++}+C_{+-}+C_{-
+}+C_{--}}\,, \label{E}
\end{equation}
where the subscript $(++)$ represents the angle set $(\theta_S, \theta_I)$ and
the $(-)$ subscript substitutes it with the orthogonal angle.

We made the $S$-parameter measurements for the low-flux case with an input pump
power of $\sim$70\,$\mu$W at a full divergence angle of 13\,mrad. A total of 16
coincidence measurements was made, each consisting of an average of 30 1-s data
sets.  We obtain S=2.739$\pm$0.119, indicating a violation of Bell's inequality
of more than 6 standard deviations.  The entanglement quality given by $S$ is
comparable to that indicated by the quantum-interference visibility
measurements.

\subsection{Quantum state tomography}

A more detailed characterization of the output state can be obtained by quantum
state tomography. One can reconstruct the density matrix, $\rho$, for a
two-photon output state through a series of projective measurements. The
coincidence counts recorded in these measurements can be used to obtain
individual density matrix elements by a maximum likelihood algorithm. We utilize
the protocol outlined in \cite{QST, altepeter}, to estimate $\rho$ from 16
projective measurements for an input pump power of 0.1\,mW at a full divergence
angle of 13\,mrad. For each projective measurement, we averaged 30 1-s
coincidence counting data points.  Coincidence data from these measurements are
used in a nonlinear least squares algorithm to estimate the bipartite density
matrix from which fidelity measures can be calculated. The real and imaginary
parts of $\rho$ are plotted in Fig.~5. This density matrix estimation yields
98.85\% fidelity for the PSI singlet state output, where fidelity is
calculated as Tr$\{\rho \rho_{\psi^{-}}\}$, with $\rho_{\psi^{-}}$ being the
singlet state density operator. Another entanglement measure, tangle, is
calculated from the measurements to be 0.9589, where a factorizable state yields
zero tangle and a maximally-entangled Bell state yields unity.  We observe that
all three types of measurements are equally effective and useful in
characterizing the PSI output state as highly entangled.

\begin{figure}[bth]\label{figure5}
\centerline{\includegraphics[width=9cm]{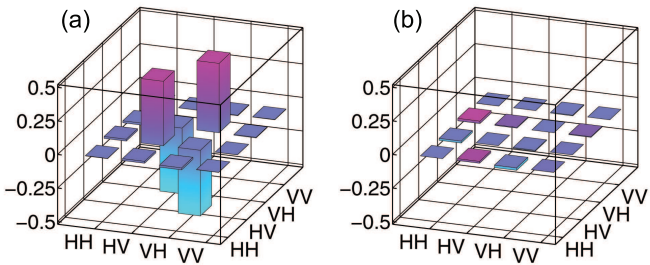}}
\caption{(Color online) Quantum state tomography of PSI singlet-state output
showing (a) real and (b) imaginary parts of the density matrix.  An optimization
routine uses a nonlinear least squares algorithm to estimate individual matrix
elements from 16 projective  measurements under different polarization analyzer
settings. Singlet-state fidelity is calculated as 98.85\%. The density matrix
yields a tangle of 0.9589.}
\end{figure}

\section{High-flux characteristics}

In the previous section the Sagnac SPDC source output has been characterized in the low-flux limit.  From the Fig.~3 measurements we have
determined that the pair generation rate was ~1\% per pulse per mW of average pump power for a full divergence angle of 13 mrad.  Availability
of over 400\,mW pump power gives us the capability to drive the SPDC process at a significant generation rate at which multiple pairs could be
produced in large numbers.  This generation capability allows one to tailor the SPDC source for a specific application. One may, for instance,
improve the entanglement source performance by increasing the pump power and reducing the divergence angle to achieve the desired generation
rate with a higher entanglement quality.  However, a higher flux output comes with its disadvantages.  The main concern is an increase of
accidental coincidences that reduces quantum-interference visibility, leading to potential errors in quantum information processing tasks. Take
entanglement-based QKD as an example, while a higher secret key rate can be obtained for a mean pair generation number $\alpha \gg 1\%$, the
maximum $\alpha$ that can be used without compromising its security depends on a number of operational system parameters \cite{ma}.  As a
result, it is highly useful to have a simple and accurate estimate of the pair generation rate under all circumstances. In this section, we will
consider the occurrence of accidental coincidences and how a pulsed source can lead to more errors than a cw source. We will also concentrate on
the effect of multiple pair generation on entanglement quality and how that effect can be utilized to measure the pair generation probability
accurately.

\subsection{Accidental coincidences}

There are three main contributions to the observation of accidental coincidences in the typical measurement setup of Fig.~1.  The first type of
contribution is due to the generation of independent multiple entangled pairs from a multimode SPDC source under strong pumping.  If the
detected coincident photon pair originates from two different signal-idler pairs, there is no polarization correlation and therefore an error
may occur. Considering the case of two-pair events, the production rate is proportional to the square of the pump power. The second type of
contribution is caused by the detection of one photon from a down-converted photon pair and a UV-induced fluorescence photon emitted by the
PPKTP crystal which we measured to have a generation rate of less than 5\% of the singles count rate.  Since the fluorescence varies linearly
with the pump power, the accidental coincidence rate is proportional to the square of the pump power.  There is also the possibility of an
accidental coincidence due to the detection of two fluorescence photons, but the probability is much lower than the case when one of the
detected photons belongs to a down-converted pair.  The third contribution comes from background photons from stray light and from detector dark
counts.  Both events are independent of the pump power and generally the background and dark counts are low enough that accidental coincidences
caused by them are negligible.

The above discussion on accidental coincidences applies to both cw and pulsed SPDC\@.  However, more accidental coincidences are observed in the
pulsed case than in the cw case for the same average pumping power.  If we assume the same SPDC setup, then the same average input power yields
the same number of down-converted pairs per second for the cw and pulsed cases.  In cw operation, the SPDC pair or fluorescence photon
generation probability within a coincidence window of duration $T_c$ is proportional to $T_c$.  The cw accidental coincidence rate resulting
from two-pair events is then given by $f_{\rm cw} = gT_c$, where $g$ is a proportionality constant.  On the other hand, for pulsed SPDC with a
repetition rate of $R_p$, the pair or fluorescence emission is localized within each pump pulse duration, whose width is typically smaller than
$T_c$, as in our setup. Therefore, the SPDC pair (or fluorescence photon) generation probability per pulse (alternatively, per coincidence
window) is proportional to $1/R_p$. The accidental coincidence probability per pulse is then proportional to $1/R_p^2$, and the accidental
coincidence rate is $f_{\rm pulsed} = g/R_p$. Note that the cw and pulsed cases have the same proportionality as long as the same type of
accidental coincidences is considered.  For the same input power, the ratio of their accidental rates is

\begin{equation}
\frac{f_{\rm cw}}{f_{\rm pulsed}}= R_p T_c\,. \label{reprate}
\end{equation}

Using a typical value for the coincidence window duration $T_c =1$\,ns, the
repetition rate of $R_p = 31.1$\,MHz for our pulsed pump implies that $f_{\rm
cw} \ll f_{\rm pulsed}$.  That is, our pulsed system is much more susceptible to
visibility degradation due to accidental coincidences compared with a cw source
with the same entangled pair generation rate.  Equivalently, the cw source can
tolerate a much higher average pump power than a pulsed source ($\sim$18$\times$
in our case) for the same amount of accidental coincidences.  This problem can
be minimized by increasing the repetition rate until it is comparable to
$1/T_c$.  Increasing the repetition rate, however, has its own problems.  For
instance, typical Si APDs are not able to handle detection rates much higher
than a  few MHz.  Also, it is not desirable to have $R_p T_c > 0.5$ because of
the need for temporal separation of the pulses.  For a pulsed source designed
for a specific application such as QKD, one must be aware of the trade-offs
between the need for a high pair generation rate and the desire for high
entanglement quality, and strike an application-specific balance for the
appropriate combination of pump powers, repetition rates, detector speeds, and
error budget.

\subsection{Multiple-pair generation}

The UV-induced fluorescence rate was measured by detuning the phase matching temperature so that only fluorescence photons were detectable and
we found that it was only a small fraction ($\sim$5\%) of the detected singles rate. Therefore multiple-pair generation is the main contributor
to accidental coincidences in our setup. The limitations due to multi-pair events have previously been observed \cite{deRiedmatten, Liang06}.
Eisenberg \emph{et al.} measured the multi-pair visibility degradation in their study of stimulated parametric emission in a multi-pass
down-conversion configuration \cite{Eisenberg}. In this section we use a simple model to quantify the effect of multiple-pair generation on
two-photon quantum interference measurements in spontaneous parametric emission in a single-pass down-conversion setup. We compare the
theoretical predictions to our experimental observations at high pump powers.

Consider an ideal pulsed SPDC source that generates singlet-state
polarization-entangled photon pairs, and there are no background or
fluorescence photons.  We assume that the free-space output is spatially
multimode and that the output is also temporally multimode because the
coincidence measurement time is large compared with the output pulse width.  The
number of entangled pairs in the output pulse is Poisson distributed with a mean
pair generation number $\alpha$:
\begin{equation}
p_n(\alpha) = \frac{e^{-\alpha}\alpha^n}{n!}\,, \label{Poisson} \\
\sum_{n=0}^{\infty} p_n(\alpha) = 1\,, \nonumber
\end{equation}
where $p_n(\alpha)$ is the probability weight for obtaining exactly $n$ pairs
per pulse.  We further assume that the $n$ entangled pairs are independent
because of the multimode nature of the output.  That is, each mode has an
occupation number of either 0 or 1.  We note that the assumption of the Poisson
distribution is not valid for single-mode outputs, but we have found that the
results below do not change significantly when we replace it with a thermal
distribution.

We focus on the effect of multiple pairs on the two-photon quantum-interference
visibility.  In calculating the visibility of Eq.~(\ref{V}), we only need to
obtain the maximum and minimum coincidence probabilities, $C_{\rm max}$ and
$C_{\rm min}$, respectively.  For singlet-state entangled output, we consider a
coincidence measurement along $H$ polarization for the signal photon and along
$V$ ($H$) polarization for the idler photon to measure $C_{\rm max}$ ($C_{\rm
min}$), and we assume ideal polarization analyzers.  Other polarization
settings (such as $A$-$D$) work just as well, as long as the polarizers are
orthogonal for measuring $C_{\rm max}$ and parallel for measuring $C_{\rm min}$.
To complete the model for the calculations, we further assume ideal
photon-number non-resolving detectors and both the signal and idler paths have a
system detection efficiency $\eta$.

The minimum and maximum coincidence probabilities are given by
\begin{eqnarray}
C_{\rm min}&=&\sum_{n=1}^{\infty} p_n(\alpha) c_{\rm min}(n)\,,  \label{Cmin} \\
C_{\rm max}&=&\sum_{n=1}^{\infty} p_n(\alpha) c_{\rm max}(n)\,,  \label{Cmax}
\end{eqnarray}
where $c_{\rm min}(n)$ and $c_{\rm max}(n)$ are the minimum and maximum
coincidence probabilities for an output with exactly $n$ pairs of entangled
photons.  Note that the summation starts from $n = 1$ because $c_{\rm
min}(0) = c_{\rm max}(0) = 0$.  For $n$ independent photon pairs, there are
$(n+1)$ different ways to arrange their polarization orientations relative to
the analyzer settings, and they follow the binomial distribution.  For example,
the $n$ signal photons may be $H$-polarized for $k$ of them (with corresponding
$V$-polarized idler photons), and $V$-polarized for the rest, or $n-k$ of them,
and it has a probability of ${n\choose k}/2^n$ with the binomial coefficient
\begin{equation}
{n\choose k} = \frac{n!}{(n-k)! k!} \label{B}
\end{equation}
for $0\le k \le n$.  For a given polarization arrangement, one requires at least
one detected $H$-polarized event at the signal detector and at least one
detected $V$-polarized event at the idler detector to yield a coincidence count
for $c_{\rm max}(n)$.  Similarly, to obtain $c_{\rm min}(n)$ it is required to
detect an $H$-polarized event at each detector.  Combining the above
requirements we obtain
\begin{eqnarray}
c_{\rm min}(n)&=& \frac{1}{2^n}\sum_{k=1}^{n-1}  {n\choose k} \left[1-(1-
\eta)^k\right] \left[1-(1-\eta)^{n-k}\right]\,, \nonumber \\
c_{\rm max}(n)&=& \frac{1}{2^n}\sum_{k=1}^{n}  {n\choose k} \left[1-(1-
\eta)^k\right]^2\,.  \label{coeff}
\end{eqnarray}
The above coefficients allow us to compute the minimum and maximum coincidence
probabilities from Eqs.~(\ref{Cmin}-\ref{Cmax}) and obtain the visibility of
Eq.~(\ref{V})  as a function of the mean pair generation probability per pulse
$\alpha$.

It is instructive to evaluate the multiple-pair effect for small $\alpha$ in
which case we ignore terms with more than two pairs of entangled photons and
simplify the Poisson distribution coefficients $p_1(\alpha) \approx
\alpha(1-\alpha)$ and $p_2(\alpha) \approx \alpha^2/2$.  In this case, we obtain
\begin{eqnarray}
C_{\rm min} &=& \frac{1}{4} \eta^2 \alpha^2\,, \\
C_{\rm max} &=& \frac{1}{2} \eta^2 \alpha \left[ 1+\frac{\alpha}{4} (2 - 4\eta +
\eta^2)\right]\,,
\end{eqnarray}
and one obtains the visibility to the first order of $\alpha$ as
\begin{equation}
V =  1- \alpha\,. \label{Vdependence}
\end{equation}
The linear dependence of the quantum-interference visibility on the mean pair
generation probability shows clearly how the entanglement quality degrades with
increasing pump powers.  It is a little surprising to see in
Eq.~(\ref{Vdependence}) that there is no dependence on the system detection
efficiency $\eta$.  A physical explanation is that for a pair of photons that
survive the system loss and are detected, one does not know if they belong to
the same entangled pair or not and therefore it should only depend on the
generation statistics and not on the system loss.  For a system with high loss
that results in a low detection rate the entanglement quality may still be
compromised by multi-pair events.

The above results can be directly compared with measurements using our setup in
Fig.~1.  The free-space output of our pulsed Sagnac source with a weakly focused
pump had many spatial modes that can be described by a Poisson distribution.
Moreover, the pumping rate was low enough that even at the highest measured
$\alpha$ value in the experiment, stimulated emission can be neglected.  We
measured quantum-interference visibilities in both $H$-$V$ and $A$-$D$ bases at
different input pump powers and at a fixed full divergence angle of 13\,mrad.
The pair generation rate per pulse $\alpha$ is inferred from singles and
coincidence measurements, which have yielded a conditional detection efficiency
of $\eta = 9.5$\% under these operating conditions. The maximum pump power was
$\sim$70\,mW which corresponds to $\alpha = 0.7$. We measured $\sim$110,000
pairs/s with $\sim$1\,MHz singles rate at this power level. We did not use
higher $\alpha$ values to avoid any possible crystal damage at elevated pump
power levels.

\begin{figure}[htb]\label{figure6}
\centerline{\includegraphics[width=8cm]{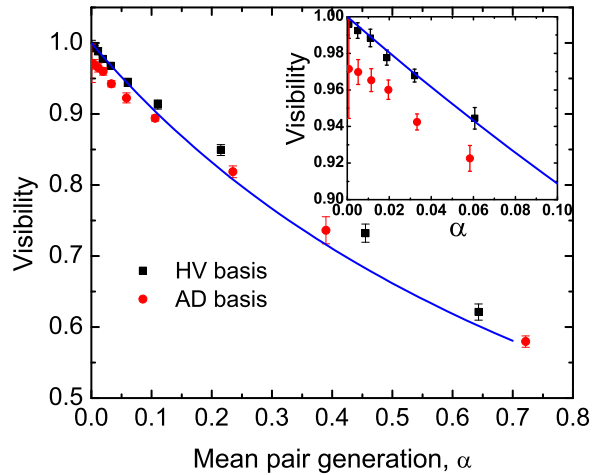}}
\caption{(Color online) Quantum-interference visibility in the $H$-$V$ (squares)
and $A$-$D$ (circles) bases as a function of mean pair generation number
$\alpha$.  The solid curve is computed from our theoretical model.  Visibilities
in the $A$-$D$ basis are lower than those in the $H$-$V$ basis due to other
factors that are unrelated to multi-pair occurrences.  Inset shows the linear
behavior of the quantum interference visibility in the low-flux regime ($\alpha
\leq$0.1).}
\end{figure}

Figure~6 plots the measured visibilities and compares them with our calculated
values.  The $H$-$V$ results match the theoretical values very well in both the
low and high $\alpha$ regimes.  In particular, at low $\alpha$, the visibility
degradation is linear with respect to the mean pair generation number. The
$A$-$D$ results also show the expected linear dependence on $\alpha$ except
there is a fixed amount of visibility loss, which is related to other factors.
We speculate that the discrepancy between the $H$-$V$ and $A$-$D$ visibilities
is due to wavefront distortion in the optical components of the Sagnac
interferometer that created partial spatial distinguishability. We have verified
that this problem did not originate from the input pump profile because we
observed no visibility improvement after we cleaned up the pump spatial mode by
coupling it into a single mode fiber. At this point, we suspect that the surface
quality of the PBS was responsible for the visibility loss.  Further testing
with a higher quality dual-wavelength PBS is needed to support this claim.
Since the $H$-$V$ measurements were made in the down-converters' natural
polarization basis, their results were not affected by these distinguishability
issues. It is useful to note that one can take advantage of the $\alpha$
dependence of $H$-$V$ visibility measurements to deduce accurately the mean pair
generation probability $\alpha$.

\section{Discussion}

We have presented a compact, pulsed, narrowband source of polarization-entangled
photons that is of interest to entanglement-based free-space QKD\@.  Pumped by a
home-built, narrowband, pulsed UV source the Sagnac geometry ensures
phase-stable generation of any desired Bell state at the degenerate wavelength
of 780.7\,nm.   In principle, the bidirectionally pumped single-crystal Sagnac
configuration eliminates spatial, spectral, and temporal distinguishability in
the interferometric combination of the two counter-propagating down-converter
outputs.  Experimentally, we observed high quality entanglement except for a
slight  quantum interference visibility loss caused probably by the spatial
distinguishability problem.  Pulsed operation allows a QKD system to easily
manage the arrival times of the photons by synchronization of the transmitter
and receiver clocks.  It also provides temporal discrimination against
background light that arrives outside of the clocked coincidence window.  The
narrowband SPDC output allows a matching narrowband spectral filter to be used
to screen out ambient light especially during daylight QKD operation.

Our pulsed Sagnac source is highly efficient and it can be pumped to have a significant pair generation per pulse, producing 0.01 pair per pulse
at an average pump power of $\sim$1\,mW at a pump repetition rate of 31.1\,MHz. We show that at high pair generation the presence of multiple
pairs degrades the two-photon entanglement quality that may not be desirable in some quantum information processing tasks.  If the pair
generation probability per pulse  is limited by multi-pair events, then the only way to increase the flux without reducing the entanglement
quality is to increase the repetition rate, up to the point where the detector speed may impose a practical limit.

It is common to evaluate the performance of a cw SPDC source by its quantum-interference visibility and its spectral brightness, which is the
number of detected pairs per second per mW of pump in a 1-nm bandwidth. However, the spectral brightness is not an appropriate performance
metric for a pulsed source.  Instead of spectral brightness, it is more useful to specify the pair generation probability per pulse and the
repetition rate of the source. With the high power pumping capability, we have observed the quantum interference visibility as a function of
pair generation rate and reached an $\alpha$ value as high as 0.7\@. The pulsed Sagnac source is capable of achieving a high enough pair
generation rate for optimal secret key generation in an entanglement-based QKD system in the presence of channel losses and non-ideal detection
efficiencies \cite{ma}. High pair generation rates have been reported for fiber-based $\chi^{(3)}$ downconversion sources \cite{Liang06,fan07}.
Our system performance compares well in terms of achievable generation efficiency with negligible amount of fluorescence and scattering effects.

In conclusion, we have developed a highly efficient SPDC source utilizing the
Sagnac configuration to generate pulsed, narrowband, polarization-entangled
photons. We have demonstrated high entanglement quality and high pair generation
rates with the Sagnac source. Under strong pumping, the pulsed source shows
entanglement quality degradation due to multi-pair events, in accordance with
our theoretical analysis.  The source can be utilized in many quantum
information processing applications including quantum key distribution.  In
addition, because of its high efficiency and compactness, the Sagnac source is
also useful for more advanced quantum enabling technologies such as an on-demand
source of single photons \cite{spod}.

\begin{acknowledgments}
This work was supported by the Disruptive Technologies Office through a National
Institute of Standards and Technology contract 60NANB5D1004.
\end{acknowledgments}


\end{document}